\documentstyle[12pt,epsfig]{article}
\begin{document}
\setlength{\baselineskip}{0.75cm}
\setlength{\parskip}{0.45cm}
\begin{titlepage}
\begin{flushright}
RAL-TR-95-071 \linebreak
November 1995
\end{flushright}
\vspace*{2.5cm}
\begin{center}
\Large{
{\bf A Rederivation of the Spin-dependent}
\vspace*{0.5cm}
{\bf Next-to-leading Order Splitting Functions}}
\vskip 4.5cm
{\large Werner Vogelsang}  \\
\vspace*{1cm}
\normalsize
Rutherford Appleton Laboratory, \\
Chilton DIDCOT, Oxon OX11 0QX, England
\end{center}
\vskip 2.5cm
\begin{center}
{\bf Abstract} \\
\end{center}
We perform a new calculation of the polarized next-to-leading order
splitting functions, using the method developed by Curci, Furmanski and
Petronzio. We confirm the results of the recent calculation by Mertig
and van Neerven.
\vskip 2cm
\end{titlepage}
\section{Introduction}
The past few years have seen much progress in our knowledge about the
nucleons' spin structure due to the experimental study of the spin
asymmetries
$A_1^N (x,Q^2)\approx g_1^N(x,Q^2)/$ $F_1^N(x,Q^2)$ ($N=p,n,d$) in
deep-inelastic scattering (DIS) with longitudinally polarized lepton beams
and nucleon targets. Previous data on $A_1^p$ by the SLAC-Yale
collaboration \cite{slac} have been succeeded by more accurate
data from [2-4], which also cover a wider range in $(x,Q^2)$,
and results on $A_1^n$ and $A_1^d$ have been published in \cite{e142} and
\cite{smcd,e143d}, respectively.

On the theoretical side, it has become possible to perform a complete
and consistent study of longitudinally polarized DIS in
next-to-leading order (NLO) of QCD, since recently  results for the
spin-dependent two-loop anomalous dimensions,
needed for the NLO evolution of polarized parton distributions,
have been calculated for the first time \cite{neerv} within the
Operator Product Expansion (OPE). A first
phenomenological NLO study has been presented in \cite{grsv}, later
followed by the analyses \cite{mel}.

The calculation of the NLO anomalous dimensions or splitting
functions is in general very
complicated. This is true in particular for the polarized case, where the
Dirac matrix $\gamma_5$ and the antisymmetric tensor $\epsilon_{
\mu\nu\rho\sigma}$ enter the calculation as projectors onto definite
helicity states of the involved particles. These (genuinely
{\em four}-dimensional) quantities lead to certain
complications when dimensional regularization - which probably represents
the only viable method of regularization in such a calculation -
is used. In fact, \cite{neerv} was recently revised since an error
related to the treatment of $\gamma_5$ was found. Although the results
of \cite{neerv} now fulfil a relation motivated from Supersymmetry -
which appears to be an important constraint - it seems necessary
to perform an independent calculation of the polarized two-loop
splitting functions to check the results of \cite{neerv}.
This is the purpose of this paper.

In the unpolarized case, two different methods have been used to
obtain the next-to-leading order splitting functions. The first
calculation \cite{flor} was performed within the
OPE. Afterwards Curci, Furmanski and Petronzio developed a technique
\cite{curci,fp}
which is as close as possible to parton model intuition since it is based
explicitly on the factorization properties of mass singularities
in the light-like axial gauge and on the generalized ladder expansion
\cite{c4}. Note that the results of \cite{fp} fulfilled the
above mentioned supersymmetric relation \cite{ant},
but were in disagreement
with the first calculation \cite{flor} and led to the detection of
an error in \cite{flor}. In this paper we exploit the
method and results of \cite{curci,fp}
to rederive the spin-dependent next-to-leading order splitting functions.
To deal with $\gamma_5$ and the $\epsilon$ tensor we use the HVBM
scheme \cite{tvbm}, which still seems to be the most consistent
prescription \cite{bon}.
Section 2 sets the framework for the calculation, details of which
are then given in section 3. In Section 4 we present our results.
\section{Framework}
In this section we outline the framework for our calculation. We mainly
focus on the new features in the polarized case; more details on the method
itself can be found in the original work \cite{curci}.

The general strategy consists of a rearrangement of the perturbative
expansion which makes explicit the factorization into a part which
does not contain any mass singularity and another one which contains all
(and only) mass singularities. Fig.1 represents the matrix element
squared for polarized virtual (space-like) photon-quark
scattering.
The blob $\Delta M$ is expanded into 2PI kernels $\Delta C_0$
and $\Delta K_0$. In the
axial gauge these 2PI kernels have been proven \cite{c5} to be finite
as long as the external legs are kept unintegrated, such that all
collinear singularities originate from the integrations over the
momenta flowing in the lines connecting the various kernels.
The generalized ladder in Fig.1 can be written as \cite{curci,foot1}
\begin{equation}
\Delta M=\Delta C_0 \left( 1+\Delta K_0+\Delta K_0^2+ ... \right)
=\Delta C_0\frac{1}{1-\Delta K_0} \equiv \Delta C_0 \Delta \Gamma_0 \:\:\: .
\end{equation}
$\Delta C_0$ and $\Delta \Gamma_0$ are then decoupled by projectors
which for the longitudinally polarized case read
($\Delta A$, $\Delta B$ being two polarized kernels):
\begin{equation}
(\Delta A) P_F (\Delta B) =
\left( \Delta A^{ij} (\not k \gamma_5)_{ij} \right)
\left[ {\rm P.P.} \right]
\left( \frac{(\gamma_5 {\not n})^{kl}}{4kn} \Delta B_{kl} \right)
\end{equation}
for polarized quarks and
\begin{equation}
(\Delta A) P_G (\Delta B) =
\left( \Delta A_{\mu\nu} \epsilon^{\mu\nu\rho\sigma}
\frac{k_{\rho} n_{\sigma}}{kn} \right) \left[ {\rm P.P.} \right]
\left( \epsilon^{\alpha\beta\gamma\delta}
\frac{k_{\gamma} n_{\delta}}{kn} \Delta B_{\alpha \beta}
\right) \:\:\: ,
\end{equation}
for polarized gluons,
putting $k^2=0$ in the part containing the kernel $\Delta A$ and taking
the pole-part (P.P.) of the projection on
kernel $\Delta B$. In eqs. (2),(3) $i,j,k,l$ are spinor
and the greek letters are Lorentz indices. $n$ is the vector to be
introduced in the axial gauge with $n^2=0$ for the light-like gauge.
Performing the factorization of mass singularities \cite{curci},
which in dimensional regularization $(d=4-2 \epsilon)$ appear
as poles in $1/\epsilon$, one ends up with the contribution
to the (partonic) deep-inelastic structure function $g_1$:
\begin{equation}
g_1 (\frac{Q^2}{\mu^2},x,\alpha_s,\frac{1}{\epsilon}) =
\Delta C (\frac{Q^2}{\mu^2},x,\alpha_s) \otimes \Delta \Gamma
(x,\alpha_s,\frac{1}{\epsilon}) \:\:\: ,
\end{equation}
where  the convolution $\otimes$ is defined as usual by
\begin{equation}
(f\otimes g)(x) \equiv \int_x^1 \frac{dz}{z} f(z)g(\frac{x}{z})  \:\:\: .
\end{equation}
In eq. (4) we have introduced the virtuality of the
photon $Q^2$, the unit of mass $\mu$ in dimensional regularization,
the Bj\o rken variable $x=Q^2/2 pq$ and the strong coupling $\alpha_s$.
Eq. (4) has a clear partonic interpretation:
$\Delta \Gamma (x,\alpha_s,\frac{1}{\epsilon})$ describes the density
of partons in the parent quark, is independent of the hard process
considered and contains all the collinear singularities (poles in
$\epsilon$), whereas $\Delta C (Q^2/\mu^2,x,\alpha_s)$ is the
(process-dependent) short-distance cross section. As was shown in
\cite{curci}, $\Delta \Gamma$ does not depend on $Q^2$, which is
a consequence of the finiteness of the kernel $\Delta K_0$ in the
axial gauge \cite{c5}, and allows for the derivation of a
'renormalization group' equation for $\Delta C$ with $\Delta \Gamma$
related to the 'anomalous dimension'. Thus $\Delta \Gamma$,
to be convoluted with bare ('unrenormalized') parton densities
which must cancel its $1/\epsilon$ poles, is equivalent to the
respective Altarelli-Parisi \cite{ap} kernels, e.g.,
\begin{equation}
\Delta \Gamma (x,\alpha_s,\frac{1}{\epsilon} )=
1-\frac{1}{\epsilon} \left[ \left(\frac{\alpha_s}{2 \pi} \right)
\Delta P_{qq}^0 (x) + \frac{1}{2} \left(  \frac{\alpha_s}{2 \pi}
\right)^2 \Delta P_{qq}^1 (x) + ... \right] + {\cal O} (\frac{1}
{\epsilon^2} )
\end{equation}
for the non-singlet (NS) case. The final NLO expression for the
(physical) spin-dependent structure function $g_1^N$ then reads:
\begin{eqnarray}
g_1 (x,Q^2) &=& \frac{1}{2} \sum_q^{N_f} e_q^2\;
\Bigg\{ \Delta q (x,Q^2)+\Delta \bar{q} (x,Q^2)+ \nonumber \\
&+& \frac{\alpha_s(Q^2)}{2\pi} \left[ \Delta C_q \otimes
\left( \Delta q +\Delta \bar{q} \right) +\frac{1}{N_f}
\Delta C_g \otimes \Delta g\right] (x,Q^2) \Bigg\}  \:\:\: ,
\end{eqnarray}
where $N_f$ is the number of active flavors, and where in the full
singlet case two short-distance cross sections $\Delta C_q$ and
$\Delta C_g$ for scattering off incoming polarized quarks or gluons,
respectively, exist.
Here, the polarized parton distributions $\Delta p \equiv
p^{\uparrow}-p^{\downarrow}$ ($p=q,g$) are to be evolved
according to the polarized Altarelli-Parisi \cite{ap} evolution equations
which to NLO read (see, e.g., \cite{v28})
\begin{eqnarray}
\frac{d}{d\ln Q^2} (\Delta q+\Delta \bar{q}-\Delta q'-
\Delta \bar{q}') &=& \frac{\alpha_s}{2\pi}
(\Delta P_{qq}+\Delta P_{q\bar{q}}) \otimes (\Delta q+\Delta \bar{q}
-\Delta q'-\Delta \bar{q}' ) \:\:\: ,  \\
\frac{d}{d\ln Q^2} (\Delta q-\Delta \bar{q}) &=& \frac{\alpha_s}{2\pi}
(\Delta P_{qq}-\Delta P_{q\bar{q}}) \otimes
(\Delta q-\Delta \bar{q})
\end{eqnarray}
for the NS quark densities and
\begin{equation}
\frac{d}{d \ln Q^2} \left( \begin{array}{c}
                           \Delta \Sigma \\ \Delta g
                           \end{array} \right) =
\frac{\alpha_s}{2\pi} \left( \begin{array}{cc}
\Delta P_{qq}+\Delta P_{q\bar{q}}+\Delta P_{qq,PS} & \Delta P_{qg} \\
\Delta P_{gq}                                      & \Delta P_{gg}
                             \end{array} \right) \otimes \left(
                             \begin{array}{c}
                             \Delta \Sigma \\ \Delta g
                             \end{array} \right)
\end{equation}
in the singlet sector,
where $\Delta \Sigma \equiv \sum_q (\Delta q+\Delta \bar{q})$ and
the argument $(x,Q^2)$ has been omitted from all parton densities.
To NLO, all splitting functions in (8-10) have the perturbative expansion
\begin{equation}
\Delta P_{ij} = \Delta P_{ij}^0 + \frac{\alpha_s}{2\pi} \Delta P_{ij}^1
\:\:\: .
\end{equation}
The entries $\Delta P_{q\bar{q}}$ and $\Delta P_{qq,PS}$ start
to be non-zero only beyond the leading order. For future reference
it is convenient to introduce the NLO combinations
\begin{equation}
\Delta P_{qq,NS}^{1,\pm} = \Delta P_{qq}^1 \pm \Delta P_{q\bar{q}}^1
\end{equation}
which according to (8,9) govern the NLO part of the evolution in the
NS sector. $\Delta P_{qq,PS}$ is called the 'pure singlet' splitting
function since it only appears in the singlet case.
\section{The Calculation}
Some representatives of the graphs to be evaluated in the calculation
of the $\Delta P_{ij}^1$ are shown in Fig.2. We do not need to calculate
the contributions from genuine two-loop graphs to the diagonal splitting
functions $\Delta P_{qq}^1$ and $\Delta P_{gg}^1$, which are
$\sim \delta (1-x)$, since these are the same as for the unpolarized case.

Before giving details of the calculation, we note that the use of
the light-like ($n^2=0$) axial gauge in practical calculations has been a
matter of debate for a long time now \cite{lrev}. The great
computational advantage and
success it has brought for, e.g., perturbative NLO QCD calculations
in DIS \cite{curci,fp,stir} or jet calculus \cite{kali} has not always
been matched by the theoretical understanding of why it worked
so well \cite{curci,bass}. The problems connected with the light-like
axial gauge are due in the first place to the presence of spurious
singularities in loop integrals coming from the vector
propagator in this gauge, which are neither
of ultraviolet nor of infrared origin. In \cite{curci,fp}
'phenomenological rules' were applied to the problem which
consisted of the subtraction of spurious poles by hand and of
using the Cauchy principle value (CPV) prescription to
deal with $1/l\cdot n$ terms (where $l$ is some momentum),
entailing renormalization 'constants' depending on the
infinite-momentum-frame (IMF) variable $x$. In view of the
complications due to the spurious poles, but also in view of the fact
that the calculations and the prescriptions in \cite{curci,fp}
led to the {\em correct} answer,
we find it very satisfactory that it turns out
to be possible in our case to infer the effective contributions of the
non-trivial \cite{foot2} virtual, in particular the vertex
correction, graphs to the polarized splitting functions from
the known results \cite{curci,fp} for the {\em un}polarized $P_{ij}^1$,
such that these contributions need not be calculated all over again.
The strategy to do this is to calculate all real emission graphs
to a given process for the unpolarized case and to subtract their
sum from the corresponding final results listed in \cite{curci,fp}.
This difference and knowledge of the pole parts (renormalization constants)
of the virtual corrections \cite{curci} then straightforwardly allow for a
determination of the result for each virtual correction in terms of a
parametrization of the most general Dirac/Lorentz structure it can
have \cite{foot}, and make a
transfer to the polarized case easily possible. In this way we avoid to
have to deal with the spurious poles again but can simply rely on the
success \cite{foot3} of the approach used in the unpolarized calculation
\cite{curci,fp}.

The calculation of the real emission graphs is
rather involved. This is true in particular for the
polarized case when using the HVBM scheme since in this method
the $d=4-2 \epsilon$ dimensional space-time is explicitly
decomposed into the usual four dimensions in which $\gamma_5$
anticommutes with the other Dirac matrices and the $-2\epsilon$
dimensional part, where it commutes. Thus the squared matrix
elements of the graphs will depend on the usual '$d$-dimensional'
scalar products like $l_1 \cdot l_2$ etc. (see Fig.2a for notation),
but also on '$(d-4)$-dimensional' ones, denoted by $\hat{l}_1\cdot
\hat{l}_2$, $\hat{k}^2$ etc. \cite{foot4}.
It is most convenient to work in the IMF parametrization of the
momenta \cite{curci} which in our case takes the form:
\begin{eqnarray}
p &=& (\:\: P\:\:,\:\:\vec{0}_{xy}\:\: ,\:\: P\:\: ,\:\:
\vec{0}_{d-4}\:\: ) \:\:\: , \nonumber \\
n &=& (\:\: \frac{pn}{2P}\:\: , \:\: \vec{0}_{xy} \:\: ,
\:\: -\frac{pn}{2P} \:\: ,\:\: \vec{0}_{d-4} \:\: )
\:\:\: , \nonumber \\
k &=& \left( \:\: xP + \frac{k^2+\tilde{k}^2}{4xP} \:\: ,
\:\: \vec{k}_T \:\: ,
\:\: xP - \frac{k^2+\tilde{k}^2}{4xP}, \:\:
\hat{\vec{k}} \:\: \right) \:\:\: , \nonumber \\
l_1 &=& (\:\: l_1^0 \:\: ,\:\: \vec{l}_1^{xy} \:\: ,
\:\: l_1^z \:\: , \:\: \hat{\vec{l}}_1 \:\: ) \:\:\: ,
\end{eqnarray}
where $x=kn/pn$ is interpreted as the IMF momentum fraction of the
incoming momentum $p$ carried by $k$, and $\tilde{k}^2 =
k_x^2+k_y^2+\hat{k}^2 \equiv k_T^2+\hat{k}^2$ is the total
transverse momentum squared
of $k$ relative to  the axis defined by $p,n$. We split
the $(d-4)$-dimensional components of $l_1$ into a part
$\hat{l}_1^{\: \parallel}$ parallel to those of $k$ and a transverse
part $\hat{l}_1^{\perp}$. According to our definitions,
only $k,l_1$, and $l_2=p-k-l_1$ possess such components.
When performing the phase space integrations one has to carefully
take into account the $(d-4)$-dimensional terms. The contribution of
each graph to $\Delta \Gamma (x,\alpha_s,\frac{1}{\epsilon})$
is given by the integration of
the projected (see eqs.(2,3)) squared matrix elements over the phase space
\begin{equation}
R \equiv \int \frac{d^d k}{(2\pi)^d} x \delta \left (x-\frac{kn}{pn}\right)
\int \frac{d^d l_1}{(2\pi)^{d-1}} \int \frac{d^d l_2}{(2\pi)^{d-1}}
(2\pi)^d \delta^{(d)} (p-k-l_1-l_2) \delta (l_1^2) \delta (l_2^2)
\end{equation}
which is conveniently written as
\begin{eqnarray}
\lefteqn{\hspace*{-0.4cm}
x^{\epsilon} (1-x)^{1-2 \epsilon} \int_{-Q^2}^{0} dk^2
(-k^2)^{1-2 \epsilon} \int_0^1 d\tilde{\kappa}
\left( \tilde{\kappa} (1-\tilde{\kappa}) \right)^{-\epsilon} \int_0^1
dw \left( w (1-w) \right)^{-\epsilon} \int_0^1 dv \left( v (1-v)
\right)^{-\frac{1}{2}-\epsilon}} \nonumber \\
&&\hspace*{-0.9cm} \times \Bigg(  (-\epsilon) \int_0^1 d\hat{\kappa}
\hat{\kappa}^{-1-\epsilon} \Bigg) \Bigg( (-\frac{1}{2}-\epsilon)
\int_0^1 d\lambda^{\perp} \left( \lambda^{\perp}\right)^{-3/2-\epsilon}
\Bigg) \Bigg( \frac{1}{\pi} \int_0^1 d\lambda^{\parallel}
\left( \lambda^{\parallel} (1-\lambda^{\parallel}) \right)^{-1/2} \Bigg)
\end{eqnarray}
where we have omitted trivial prefactors and defined
\begin{eqnarray}
\hat{k}^2 &=& -k^2 (1-x) \hat{\kappa} \tilde{\kappa}
\:\:\: , \nonumber \\
\tilde{k}^2 &=& -k^2 (1-x) \tilde{\kappa} \:\:\: , \nonumber \\
l_1^0 + l_1^z &=& 2 P (1-x) w
=2 P \frac{l_1 n}{pn} \:\:\: , \nonumber \\
(l_1^0)^2-(l_1^z)^2 &=& c_1^2 + v (c_2^2-c_1^2)
= \frac{1}{P} (l_1^0+l_1^z) (l_1 p) \:\:\: , \nonumber \\
\hat{l}_1^{\: \parallel} &=& \lambda_1 + \lambda^{\parallel}
(\lambda_2 - \lambda_1) \:\:\: , \nonumber \\
(\hat{l}_1^{\perp})^2 &=& v (1-v) \left( c_1+c_2 \right)^2
\lambda^{\perp}
\end{eqnarray}
with
\begin{eqnarray*}
c_{1,2} &\equiv& \sqrt{\frac{-k^2 (1-x) w}{x}} \Bigg[
\sqrt{(1-w) (1-\tilde{\kappa})}\mp \sqrt{x w \tilde{\kappa}} \Bigg]
\:\:\: ,  \\
\lambda_{1,2} &=& -\frac{1}{2} \frac{\hat{\kappa}}{\hat{k}w}
\left( (l_1^0)^2-(l_1^z)^2 -c_1 c_2 \right)
\mp (c_1 + c_2 )\sqrt{(1-\hat{\kappa})(1-\lambda^{\perp})v (1-v)}
\:\:\: .
\end{eqnarray*}
Note that the last two integrals in (15) are all unity if no dependence
on $(d-4)$-dimensional scalar products occurs, which of course is
always the case in the unpolarized situation. If present, such
$(d-4)$-dimensional terms only give contributions proportional to
$\epsilon$ after the last three integrals in (15) have been performed.

Following \cite{curci,fp} we will regularize infrared divergencies
appearing at $l_1 n \rightarrow 0$ or $l_2 n \rightarrow 0$
($w\rightarrow 0$ or $w\rightarrow 1$), which are typical of
the axial gauge, by the CPV prescription (see above):
\begin{equation}
\frac{1}{l n}\rightarrow \frac{l n}{(l n)^2+\delta^2 (pn)^2}
\:\:\: .
\end{equation}
All the resulting divergencies of this type can then be
transformed into the basic integrals
\begin{equation}
I_i \equiv \int_0^1 dy \frac{y \ln^i y}{y^2+\delta^2}
\:\:\:\:\:\:\: (i=0,1) \:\:\: .
\end{equation}
$I_0$, $I_1$ have to cancel out in the final answer which,
as well as the finiteness of each graph in the axial gauge
\cite{curci,c5} before the final $k^2$ integration
is performed, helps when extracting the contribution of the virtual
corrections by the strategy outlined above.
As mentioned above, the renormalization constants depend
on the IMF fractions $x$ or $1-x$ when the CPV prescription is used
\cite{curci}. We finally note that whenever considering a genuine ladder
graph with two parallel rungs, subtraction of the 'doubly collinear'
graph (see Fig.2) is required within the method of \cite{curci}.
The result for this is given by convoluting the $n$-dimensional
leading order splitting function standing for the upper part
with the four-dimensional one representing the lower part of the
diagram, and including a factor $(1-x)^{-\epsilon}$ from
phase space in the convolution.
In $4-2 \epsilon$ dimensions the polarized LO splitting functions
read for $x \neq 1$ in the HVBM scheme \cite{gv}:
\begin{eqnarray}
\Delta P_{qq}^0 (x,\epsilon) &=& C_F \left( \frac{1+x^2}{1-x} +
3 \epsilon (1-x) \right) \:\:\: , \nonumber \\
\Delta P_{qg}^0 (x,\epsilon) &=& 2 T_R N_f \left( 2 x-1 -
2 \epsilon (1-x) \right) \:\:\: , \nonumber \\
\Delta P_{gq}^0 (x,\epsilon) &=& C_F \left( 2-x +2 \epsilon (1-x) \right)
\:\:\: , \nonumber \\
\Delta P_{gg}^0 (x,\epsilon) &=& 2 C_A \left( \frac{1}{1-x}-2 x+1 +
2 \epsilon (1-x) \right) \:\:\: ,
\end{eqnarray}
where $C_F=4/3$, $C_A=3$, $T_R=1/2$ and $N_f$ is the number of
active flavors.
\section{Results}
In the normalization of \cite{curci,fp} our
$\protect{\overline{{\mbox{\rm MS}}}}$ results read:
\begin{eqnarray}
\Delta P_{qq,NS}^{1,\pm} (x) &=& P_{qq,NS}^{1,\mp} (x) - 2 \beta_0
C_F (1-x) \:\:\: , \\
\Delta P_{qq,PS}^1 (x) &=& \Delta P_{qq,PS}^{(\protect{\cite{neerv}})}
(x) \:\:\: , \nonumber \\
\Delta P_{qg}^1 (x) &=& \Delta P_{qg}^{(\protect{\cite{neerv}})}
(x) + 4 C_F (1-x) \otimes \Delta P_{qg}^0 (x) \:\:\: , \nonumber \\
\Delta P_{gq}^1 (x) &=& \Delta P_{gq}^{(\protect{\cite{neerv}})}
(x) - 4 C_F (1-x) \otimes \Delta P_{gq}^0 (x) \:\:\: , \nonumber \\
\Delta P_{gg}^1 (x) &=& \Delta P_{gg}^{(\protect{\cite{neerv}})} (x)
\:\:\: ,
\end{eqnarray}
where $\beta_0=11 C_A/3-4 T_R N_f/3$ and $\Delta P_{ij}^0
(x) \equiv \Delta P_{ij}^0 (x,0)$ (see eq. (19)).
The $\Delta P_{ij}^{(\protect{\cite{neerv}})}$ \cite{foot5}
are the results of \cite{neerv}, and $P_{qq,NS}^{1,\pm}$ can be
found in \cite{curci}.
As was already discussed in \cite{grsv,ms} and indicated in eq. (20),
the '$+$' and '$-$' combinations of the NS splitting functions as
defined in (12) interchange their
role in the polarized case, such that, according to eqs.(8,12,20),
the combination $\Delta P_{qq,NS}^{1,+}=P_{qq}^1 {\bf -}
P_{q\bar{q}}^1 - 2\beta_0 C_F (1-x)$ would govern the $Q^2$-evolution
of, e.g., the polarized NS quark combination
$$\Delta A_3 (x,Q^2) = \left(
\Delta u +\Delta \bar{u}- \Delta d -\Delta \bar{d} \right) (x,Q^2)
\:\:\: .$$
Since the first moment (integral) of the latter corresponds
to the nucleon matrix element of the NS axial vector current
$\bar{q} \gamma^{\mu} \gamma_5 \lambda_3 q$ which is conserved,
it has to be $Q^2$-independent \cite{kod}. Keeping in mind that
the first moment of the unpolarized $P_{qq}^1-P_{q\bar{q}}^1$
vanishes already due to fermion number conservation
\cite{curci}, it becomes obvious that the additional term $-2 \beta_0
C_F (1-x)$ in (20) spoils the $Q^2$ independence of the first
moment of $\Delta A_3 (x,Q^2)$.
It is therefore necessary to perform a factorization scheme transformation
to the results in (20,21) in order to remove this additional term
which, as pointed out in \cite{alex,gv,ms}, is typical of the HVBM scheme
with its not fully anticommuting $\gamma_5$ and trivially would not be
present in a scheme with a fully anticommuting $\gamma_5$ since
then the two $\gamma_5$ matrices appearing in the relevant graphs
could be removed by anticommuting them towards each other
and using $\gamma_5^2=1$ (cf. Fig2a). We note that all these observations
were already made in the original calculation of \cite{neerv}
in the OPE where, however, the removal of the additional term
$\sim (1-x)$ corresponds to a finite renormalization rather than
a factorization scheme transformation. It is nice to recover
this analogy between our results and those of \cite{neerv,foot6}.
The factorization scheme transformation for $\Delta P_{qq,NS}^{1,\pm}$
also affects the singlet sector since, according to eq. (10),
$\Delta P_{qq,NS}^{1,+}+\Delta P_{qq,PS}^1$ occurs in the evolution
of the NLO quark singlet $\Delta \Sigma$.
The transformation reads in general (see, e.g., \cite{neerv,gr}):
\begin{eqnarray}
\Delta P_{qq,NS}^{1,\pm} &=& \Delta \tilde{P}_{qq,NS}^{1,\pm}
-2 \beta_0 z_{qq} \:\:\: , \nonumber \\
\Delta P_{qq,PS}^1 &=& \Delta \tilde{P}_{qq,PS}^1
\:\:\: , \nonumber \\
\Delta P_{qg}^1 &=& \Delta \tilde{P}_{qg}^1
+ 4 z_{qq} \otimes \Delta P_{qg}^0 \:\:\: , \nonumber \\
\Delta P_{gq}^1 &=& \Delta \tilde{P}_{gq}^1
- 4 z_{qq} \otimes \Delta P_{gq}^0 \:\:\: , \nonumber \\
\Delta P_{gg}^1 &=& \Delta \tilde{P}_{gg}^1  \:\:\: ,
\end{eqnarray}
where the $\Delta \tilde{P}_{ij}^1$ now are the NLO splitting
functions on the left-hand-sides of eqs. (20,21) and the $\Delta P_{ij}^1$
are the {\em new} splitting functions after the scheme transformation.
One immediately sees that the choice
\begin{equation}
z_{qq} = - C_F (1-x)
\end{equation}
leads to $\Delta P_{qq,NS}^{1,\pm}=P_{qq,NS}^{1,\mp}$ and
thus now yields the required vanishing of the first moment of
$\Delta P_{qq,NS}^{1,+}$. Even more, the transformation (22,23)
removes {\em all} additional terms on the right-hand-side
of (20,21) simultaneously, bringing our final result into
complete agreement with the revised one of \cite{neerv}.
We finally note that the above factorization scheme transformation
also changes the quark short distance cross section (coefficient
function) $\Delta C_q$ in (7), since the combination
$$\Delta C_q - \frac{2 \Delta P_{qq,NS}^{1,\pm}}{\beta_0}$$
must be independent of the choice of the factorization scheme
convention \cite{gr}. As was shown in \cite{alex,arg,ms},
only {\em after} the transformation (22,23) takes $\Delta C_q$ in the
$\overline{{\mbox{\rm MS}}}$ scheme the form
\begin{eqnarray}
\Delta C_q(x) &=& C_F \Bigg[(1+x^2) \left(\frac{\ln (1-x)}{1-x}
\right)_{\!\!+}
-\frac{3}{2} \frac{1}{(1-x)_+} -\frac{1+x^2}{1-x} \ln x +
\nonumber \\
& &  +\, 2 + x - \left(\frac{9}{2}+\frac{\pi^2}{3}\right) \delta (1-x)
\Bigg]  \:\:\: ,
\end{eqnarray}
i.e., becomes the previously calculated \cite{kod2}
${\cal O}(\alpha_s)$ quark-correction to $g_1$ giving rise to, e.g.,
the correct first order correction $1-\alpha_s/\pi$ to the
Bj\o rken sum rule. In eq. (24),
$$\int_0^1 dz f(z) \left( g(z) \right)_+
\equiv \int_0^1 dz \left( f(z)-f(1) \right) g(z) \:\:\: . $$
For completeness we note that the NLO $\overline{{\mbox{\rm MS}}}$ gluonic
short distance cross section $\Delta C_g$ in eq. (7) remains
unaffected by the transformation (22,23) and reads (see, e.g.,
\cite{neerv})
\begin{equation}
\Delta C_g(x) = 2 T_R N_f \left[ (2x-1) \left(\ln \frac{1-x}{x}-1\right)+
2(1-x)\right] \:\:\: .
\end{equation}

Our {\em final} results for the polarized $\overline{{\mbox{\rm MS}}}$
NLO splitting functions are given by:
\begin{eqnarray}
\Delta P_{qq,NS}^{1,\pm} &=& P_{qq,NS}^{1,\mp}  \:\:\: , \\
\Delta P_{qq,PS}^1 (x) &=& C_F T_R N_f \Bigg[ 2 \left( 1 - x \right)
-2 \left( 1 - 3 x \right) \ln x  - 2 \left( 1 + x \right)
\ln^2 x \Bigg]  \:\:\: , \\
\Delta P_{qg}^1 (x) &=& C_F T_R N_f \Bigg[ -22 + 27 x - 9 \ln x +
     8 \left( 1 - x \right) \ln (1-x) \nonumber \\
&& + \frac{1}{2}
     \delta p_{qg}(x) \left( 4 \ln^2 (1-x) -
      8 \ln (1-x) \ln x + 2 \ln^2 x - 8 \zeta(2) \right)
     \Bigg] \nonumber \\
&&+ C_A T_R N_f \Bigg[ 2 \left( 12 - 11 x \right)  -
     8 \left( 1 - x \right) \ln (1-x) + 2 \left( 1 + 8 x \right) \ln x
     \nonumber \\
&& \left. - 2 \left( \ln^2 (1-x) - \zeta(2) \right) \delta p_{qg}(x)
- \left( 2 I_x - 3 \ln^2 x \right)  \delta p_{qg}(-x) \Bigg] \right.
\:\:\: , \\
\Delta P_{gq}^1 (x) &=& C_F T_R N_f \left[ -{{4}\over 9} (x+4)
- \frac{4}{3} \delta p_{gq}(x) \ln (1-x) \right] \nonumber
\\ &&+ C_F^2
   \left[ - \frac{1}{2} -
   \frac{1}{2} \left( 4 - x \right) \ln x - \delta p_{gq}(-x) \ln (1-x)
\right. \nonumber \\
&& \left. + \left( - 4 - \ln^2 (1-x) + \frac{1}{2} \ln^2 x \right)
    \delta p_{gq}(x) \right]  \\
&& + C_A C_F \left[ \left( 4 - 13 x \right) \ln x +
   \frac{1}{3} \left( 10 + x \right) \ln (1-x) +
  \frac{1}{9} \left( 41 + 35 x \right) \right. \nonumber \\
&& + \left. \frac{1}{2} \left( -2 I_x + 3 \ln^2 x \right)
     \delta p_{gq}(-x) +
  \left( \ln^2 (1-x) - 2 \ln (1-x) \ln x - \zeta(2) \right)
  \delta p_{gq}(x) \right]  \nonumber \\
\Delta P_{gg}^1 (x) &=&
- C_A T_R N_f \left[ 4 \left( 1 - x \right)  + \frac{4}{3}
     \left( 1 + x \right) \ln x + \frac{20}{9} \delta p_{gg}(x)
     +\frac{4}{3} \delta (1-x) \right] \nonumber \\
&& - C_F T_R N_f
  \Bigg[ 10 \left( 1 - x \right)  + 2 \left( 5 - x \right) \ln x  +
     2 \left( 1 + x \right) \ln^2 x
     +\delta (1-x) \Bigg]  \nonumber \\
&& + C_A^2 \Bigg[ \frac{1}{3} \left( 29 - 67 x \right) \ln x -
     \frac{19}{2} \left( 1 - x \right) +
     4 \left( 1 + x \right) \ln^2 x -2 I_x \delta p_{gg}(-x)
\nonumber \\
&& + \left( {{67}\over 9} - 4 \ln (1-x) \ln x +
        \ln^2 x - 2 \zeta(2) \right) \delta p_{gg}(x)
 +\left( 3 \zeta(3) +\frac{8}{3} \right) \delta (1-x) \Bigg]
\end{eqnarray}
where, as mentioned above, the unpolarized NS pieces
$P_{qq,NS}^{1,\pm}$ can be found in \cite{curci} and
\cite{footm}
\begin{eqnarray}
\delta p_{qg} (x) &\equiv& 2 x-1  \:\:\: , \nonumber \\
\delta p_{gq} (x) &\equiv& 2 - x  \:\:\: , \nonumber \\
\delta p_{gg} (x) &\equiv& \frac{1}{(1-x)_+} - 2 x + 1   \:\:\: .
\end{eqnarray}
Furthermore we have in eqs. (26-30) $\zeta(2)=\pi^2/6$,
$\zeta(3)\approx 1.202057$ and
$$I_x\equiv \int_{x/(1+x)}^{1/(1+x)} \frac{dz}{z} \ln \Bigg(
\frac{1-z}{z} \Bigg) \:\:\: .$$
For relating our results to those of \cite{neerv} the relation
$$I_x = -2 Li_2 (-x)-2 \ln x \ln (1+x)+\frac{1}{2} \ln^2 x-\zeta (2)$$
is needed, where $Li_2 (x)$ is the Dilogarithm \cite{dd}. In
(30), the contributions $\sim \delta (1-x)$ to $\Delta P_{gg}^1$
are the same as those for the unpolarized $P_{gg}^1$
\cite{wada}; they lead to satisfaction of the constraint \cite{ar}
$$\int_0^1 dx \Delta P_{gg}^1 (x) = \frac{\beta_1}{4} \equiv
\frac{17}{6} C_A^2 - C_F T_R N_f-\frac{5}{3} C_A T_R N_f \:\:\: ,$$
valid in the $\overline{{\mbox{\rm MS}}}$ scheme.

In conclusion, our calculation, which was based on the approach
of \cite{curci} and on using the HVBM \cite{tvbm} prescription
for $\gamma_5$, has confirmed the recent results
of \cite{neerv} for the spin-dependent two-loop splitting functions
$\Delta P_{ij}^1 (x)$. Our results also once more demonstrate the
usefulness and applicability of the method of \cite{curci} and the
light-like axial gauge in perturbative QCD calculations.

\section*{Figure Captions}
\begin{description}
\item[Fig.1] The matrix element squared for polarized photon-quark
interaction, its expansion in terms of 2PI kernels $\Delta C_0$
and $\Delta K_0$, and its final factorized form.
\item[Fig2.] Some representative Feynman graphs to be evaluated in the
calculation of {\bf a:} $(\Delta) P_{qq}^1$, {\bf b:}
$(\Delta) P_{q\bar{q}}^1$, $(\Delta) P_{qq,PS}^1$, {\bf c:}
$(\Delta) P_{qg}^1$, {\bf d:} $(\Delta) P_{gq}^1$, {\bf e,f:}
$(\Delta) P_{gg}^1$. Subtraction of 'doubly collinear' graphs is indicated.
\end{description}
\newpage
\pagestyle{empty}
\vspace*{4.5cm}
\hspace*{-0.3cm}
\epsfig{file=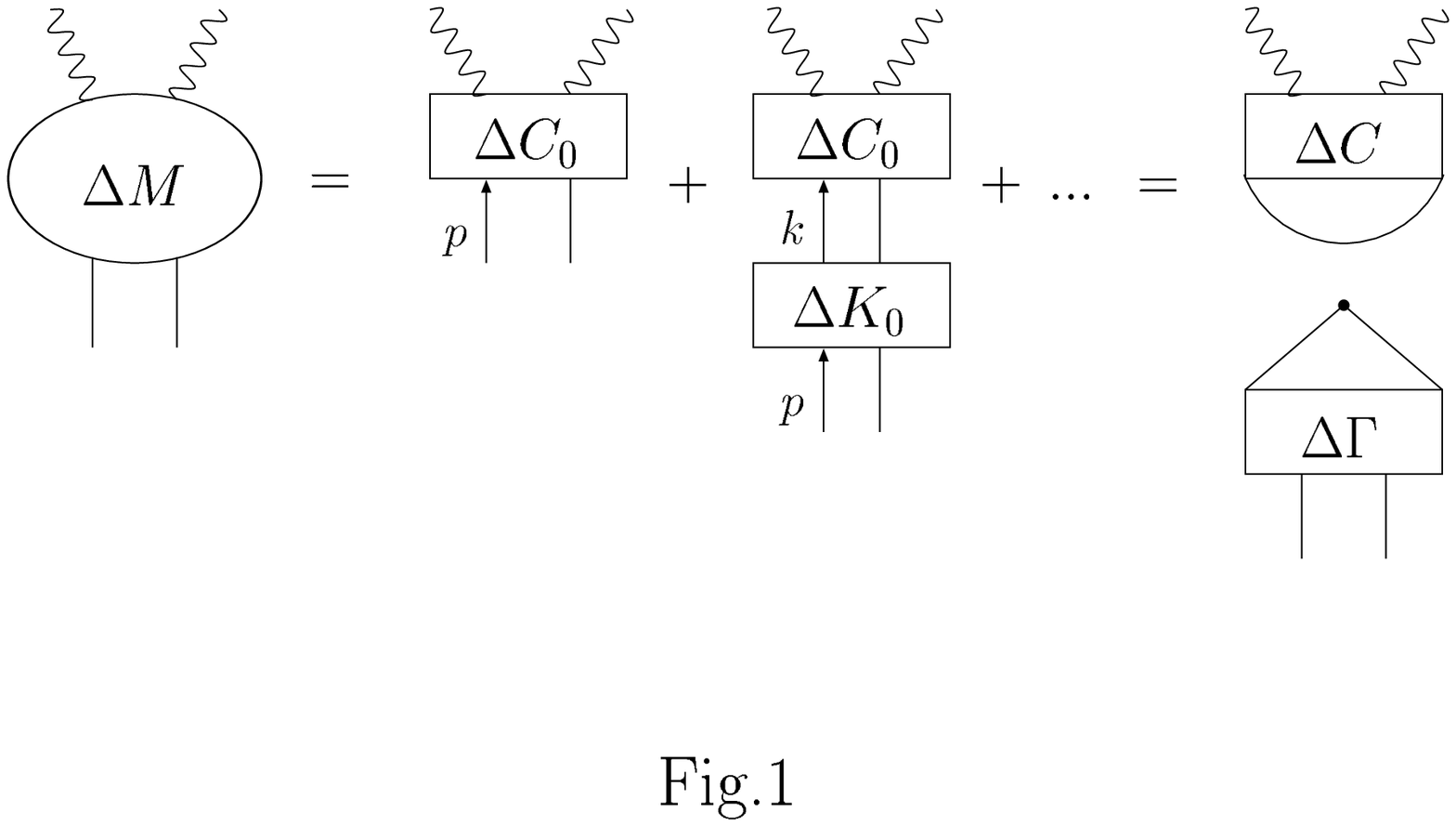,width=16cm,angle=0}
\newpage
\vspace*{-3cm}
\hspace*{-1.4cm}
\epsfig{file=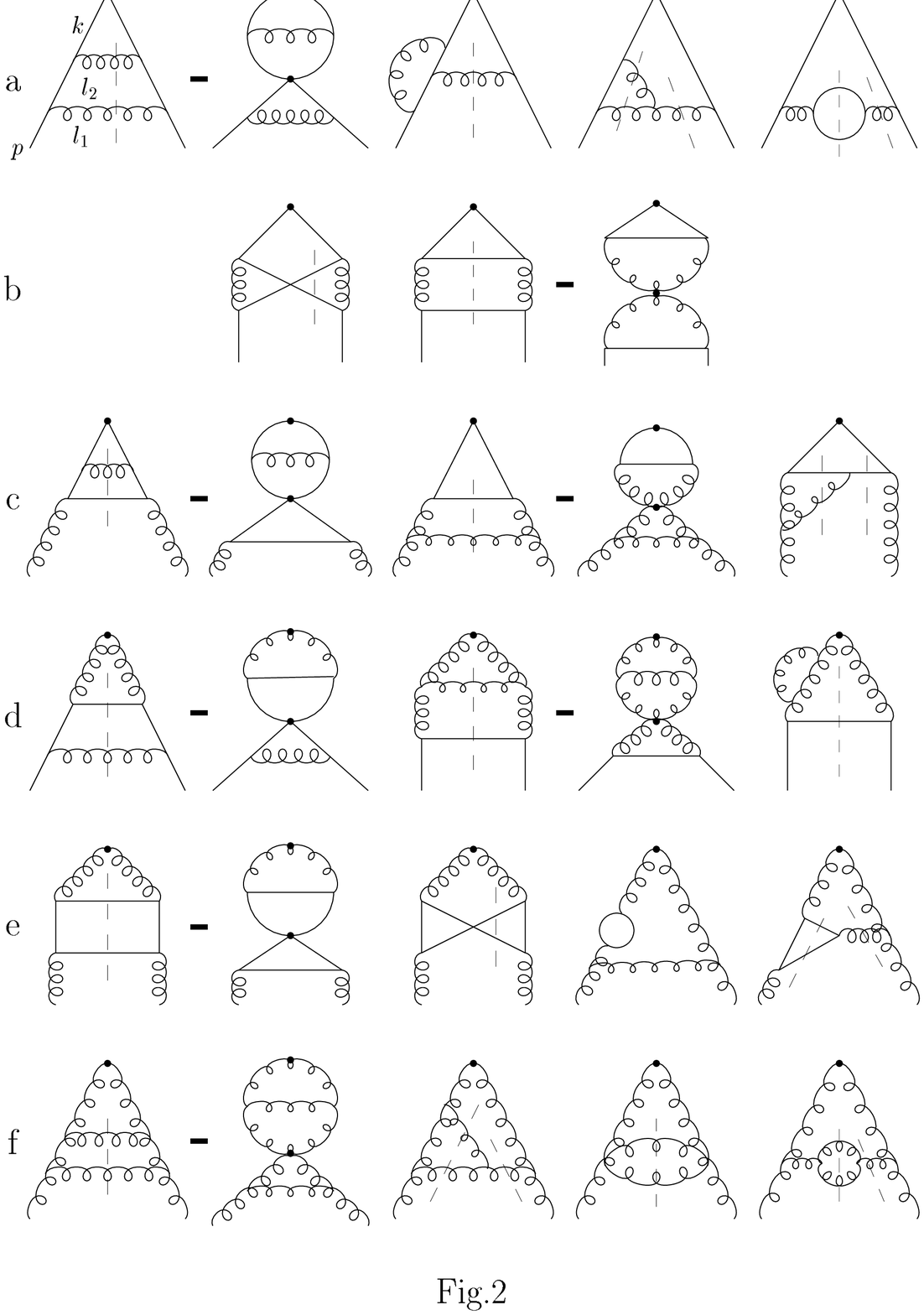,width=18cm,angle=0}

\begin{thebibliography}{99}
\bibitem{foot2} By these we mean virtual corrections which do not
enter just via their (known) renormalization constants or
as pure quark loops (the last graph in Fig.2a is an example for
both cases).
\bibitem{foot3} We note that it seems desirable for the future to put
the results of \protect{\cite{curci,fp}} (and ours) on a somewhat
field-theoretically firmer basis. First steps in this direction
have been taken in ref. \protect{\cite{bass}}.
\bibitem{foot5} In this normalization the results of
\protect{\cite{neerv}} have to be divided by a factor 8.
\bibitem{foot6} We note that in an original version of
\protect{\cite{neerv}}
the subtraction of the terms $\sim (1-x)$ was performed incorrectly
which was responsible for the error mentioned in the introduction.
\bibitem{footm} Needless to say that for $\delta p_{gg} (-x)$ the
'$+$'-prescription has to be omitted.
\end{thebibliography}
\end{document}